\documentclass[color]{article}
\usepackage{graphicx}
\usepackage[usenames,dvipsnames]{xcolor}
\usepackage{float}

\begin{document}

\centerline{\large \bf Dynamic Ising Model:}
\vskip 5pt
\centerline{\large \bf Reconstruction of Evolutionary Trees}
\vskip5pt

\centerline{P.M.C. de Oliveira}

\vskip5pt\noindent
Instituto de F\'{\i}sica, Universidade Federal Fluminense\\
Av. Litor\^{a}nea s/n, Boa Viagem, Niter\'{o}i 24210-340, RJ, Brazil\\
and National Institute of Science and Technology for Complex Systems

\noindent
e-mail address: pmco@if.uff.br

\bigskip

\begin{abstract}

	An evolutionary tree is a cascade of bifurcations starting from a single common root, generating a growing set of daughter species as time goes by. Species here is a general denomination for biological species, spoken languages or any other entity evolving through heredity. From the $N$ currently alive species within a clade, distances are measured through pairwise comparisons made by geneticists, linguists, etc. The larger is such a distance for a pair of species, the older is their last common ancestor. The aim is to reconstruct the past unknown bifurcations, i.e. the whole clade, from the knowledge of the $N(N-1)/2$ quoted distances taken for granted. A mechanical method is presented, and its applicability discussed.

\end{abstract}


\vskip50pt

\section{Introduction}

The famous Ising model deals with discrete dynamic variables $S_i = \pm 1$ for a set of ``spins'' $i = 1, 2, 3 \dots N$. Its ``energy'' is

$$ E = \sum_{\rm links} J_{ij} S_i S_j $$

\noindent where $J_{ij} = J_{ji}$ are known coupling constants and the quotes mean the absence of a proper dynamics. One cannot take gradients of this ``energy'', as in Newtonian dynamics for instance, simply because its dynamic variables are discrete. In the absence of a proper dynamics, one normally resorts to artificial ones borrowed from equilibrium statistical mechanics (Metropolis, etc). The term ``spins'' also deserves quotes, because besides the original magnetic interpretation there are a lot of distinct applications, since the also famous lattice-gas model where the values $\pm 1$ represent the presence or absence of a ``molecule'', until the modern agent-based social models where $\pm 1$ represent alternative individual opinions, votes, etc. The number of applications is huge, the Ising model is surely by far the most used statistical mechanics model in History.

	In this text, we introduce a very simple continuous version of the Ising model, by replacing the discrete variables $S_i$ by real values $x_i$. The (now unquoted) energy reads

$$ E = \frac{1}{2} \sum_{ij} J_{ij} (x_i - x_j)^2 \eqno(1)$$

\noindent The pair energy in Equation (1) can be divided in two terms $J_{ij} (x_i^2 + x_j^2)$ and $-\, J_{ij} x_i x_j$. The first term can be interpreted as an external field acting in each particle separately. The second one corresponds to their effective interaction, a generalization of the standard Ising model $-\, J_{ij} S_i S_j$. The difference is the continuous character of the dynamical variables $x_i$, which allows one to adopt Newtonian dynamics. This behavior may open the door for a lot of future applications in distinct systems. For instance, by averaging many different random sets of $J_{ij}$, this model is a generalization of the Sherrigton-Kirkpatrick spin-glass model, which can thus be studied with the help of Newtonian dynamics.

	Here, we restrict ourselves to a very particular application, the reconstruction of evolutionary trees, by following the movement of particles along an axis, for a fixed set of $J_{ij}$: it is a simple mechanical problem.

\section{Evolutionary trees}

	Natural experiments \cite{Diamond} are those where the experimenter cannot manipulate the object of study. Only comparisons can be made. It is a recent field of research allowing quantitative studies of historical evolutions. Figure 1 exemplifies a clade. At left, the traditional cladogram showing the successive speciations. This kind of draw is familiar to geneticists, linguists, etc. A good historical description entitled {\sl Trees before and after Darwin} was recently published \cite{Tassy}. At right, on Figure 1, we add the not so familiar concept of ultrametric distances on which our analysis is based. A direct measure of such a distance demands scarce fossil data. However, researchers perform indirect measurements of such pairwise distances by comparing features of currently alive species. How to perform these measurements is a vast field of research, out of the current scope. A good review can be found in \cite{phylogeny} and references therein. Human evolution can be traced back with genetic or linguistic measurements \cite{Cavalli-Sforza}. A further, still incomplete list of works is shown in \cite{genetic} for genetics, and \cite{linguistic} for linguistics. Here, we simply suppose the pairwise distances of a given clade were measured among its $N$ current alive species, i.e. a set of $N(N-1)/2$ positive numbers taken for granted. The purpose is to reconstruct the whole tree from these data.

\begin{figure}[H]
 \vskip-50pt
 \begin{center}
 \input{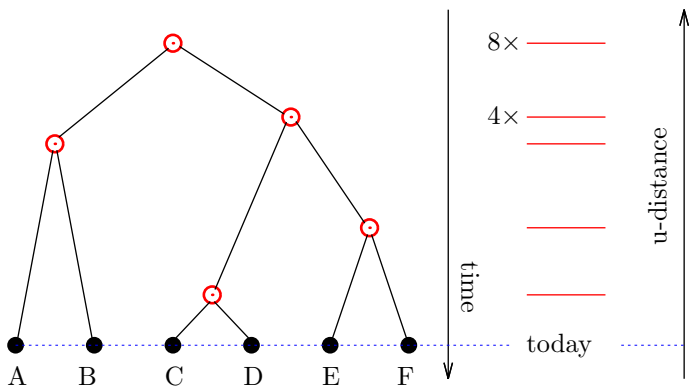}
 \end{center}
 \vskip-30pt
 \caption{Schematic clade: Closed circles represent the 6 known currently alive species, open circles their past ancestors. The ultrametric distance between two alive species is the time counted from today back to their last common ancestor: 15 distances $D_{ij}$ are shown in the spectrum at right. The uppermost level is 8-fold degenerate, i.e. the same distance appears 8 times. It corresponds to the root, the single original species. The second level is also 4-fold degenerate. The whole clade (family) is divided in two sub-clades (genera). Currently alive species A and B form one genus, whereas C, D, E and F form a second one.}

\label{dynising5}
\end{figure}

	In short, by knowing the distances exemplified at right in Figure 1, the problem is to draw the corresponding tree at left. A mechanical solution follows.

\section{The method}

	Each currently alive species is associated to a unitary mass particle moving along a $X$ axis. Particles are transparent, they can pass through each other. All $N$ particles are initially released at the origin $x=0$ with random velocities (zero sum, keeping the center of mass at rest). Particles interact through the energy (1), where the sum runs
over all pairs ($i,j = 1, 2, \dots N$), with coupling constants given by

$$ J_{ij} = D - D_{ij} \eqno(2)$$

\noindent where $D$ is an adjustable parameter (we will get rid of it soon). $D_{ij}$ are the quoted distances. The movement follows Newton's law, the accelerations

$$ \ddot{x}_i = -\sum_j J_{ij} (x_i-x_j) \eqno(3)$$

\noindent form a set of $N$ linear, second order differential equations which can be solved by diagonalizing its corresponding $N \times N$ secular matrix. Before that, let's foresee the movement.

	Take $D$ in between the two uppermost levels, Figure 1. If alive species $i$ and $j$ belong to the same genus, the coupling constant $J_{ij}$ is positive (attraction). Otherwise, $J_{ij}$ is negative (repulsion). The two genera repel each other, while attraction holds inside each genus. Therefore, $n$ (or $N-n$) particles belonging to one (or the other) genus remain clustered running away towards one (the other) sense along the $X$ axis. The eventual partition defines two genera. The same process is repeated within each just discovered genus, and so on, reconstructing the whole clade.

\section{Matrix approach}

	The secular matrix of Equation (3) can be divided as

$${\bf S} - N\hskip-2ptD\, {\bf I} + D\, {\bf G} \eqno(4)$$

\noindent where ${\bf I}$ is the identity and ${\bf G}$ is a $N \times N$ matrix with all entries $G_{ij}=1$. Matrix

$$
 {\bf S} = \left(\begin{array}{cccc}
 \sum D_{1j} & -D_{12} & -D_{13} & \dots\\
 -D_{21} & \sum D_{2j} & -D_{23} & \dots\\
 -D_{31} & -D_{32} & \sum D_{3j} & \dots\\
 \dots & \dots & \dots & \dots 
\end{array} \right)
\eqno(5)
$$

\noindent does not depend on (hereafter discarded) $D$, only on the measured distances $D_{ij}$. Their $N$ eigenvectors completely define the movement. Among them, two deserve particular comments, two next paragraphs.

	The Goldstone eigenvector has $N$ unitary entries $(1, 1, 1 \dots 1, 1)$, whose eigenvalue is always null. Interaction energy (1) presents only internal forces between the $N$ particles themselves, thus the global center of mass remains at rest at the origin, i.e. $x_1 + x_2 + x_3 \dots x_{N-1} + x_N = 0$. Furthermore, any other eigenvector $(a_1, a_2, a_3 \dots a_{N-1}, a_N)$, is orthogonal to this always-present Goldstone, i.e. $a_1 + a_2 + a_3 \dots a_{N-1} + a_N = 0$. In other words, any eigenvector besides the Goldstone is a series of positive and negative entries with zero sum. Matrix ${\bf G}$ nullifies all these further eigenvectors.

	Among them, the eigenvector with highest eigenvalue presents $n$ positive entries for one genus, $N-n$ negative entries for the other genus,
thus solving our problem. Let's call it the partition eigenvector. For the simple clade shown at left in Figure 2, for instance, the partition eigenvector is $(2, -1, -1)$. For the largest clade at right, it is $(N-n, N-n, N-n \dots -n, -n)$.

\begin{figure}[H]
 \vskip-70pt
 \begin{center}
 \input{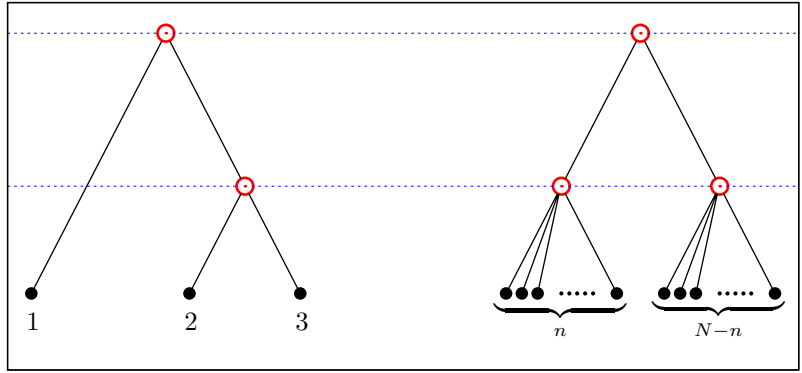}
 \end{center}
 \vskip-30pt
 \caption{Some analytically solved trees.}

\label{analytic}
\end{figure}

	 The other $N-2$ eigenvectors are unimportant, governing only the internal movement within each already-separated genus.	In short, given some $D_{ij}$ dataset, the only task is to compose matrix (5), finding its eigenvector with highest eigenvalue. The signs of its entries describe the correct partition.

\section{Data fluctuations}

	However, measured distances suffer from drift fluctuations imposed by the past evolution randomness. Therefore, they do not reflect the exact degeneracies. Each degenerate level becomes a band of neighboring levels, no longer degenerate. In Figure 1, the highest and second highest levels would be represented by two bands with 8 and 4 separated levels, respectively. Ultrametricity is lost.

	Once a given bifurcation was successfully reconstructed, the corresponding broken degeneracy can be restored as follows. Let's $n_1$ and $n_2$ be the number of species in each branch of the quoted bifurcation. Then, the $n_1 n_2$ corresponding distances displayed in the measured dataset represent indeed different measurements of the same single value, namely the real ultrametric distance. It is, then, better represented by the average over measurements: one replaces all $n_1 n_2$ distances in the $D_{ij}$ original dataset by their average, restoring the corresponding degeneracy. (Besides the average, dispersion serves to estimate the age uncertainty.)

	The band widths increase with the total evolutionary time, due to accumulated random drift. If the clade under study is too old, these bands tend to overlap over each other, and the model may fail beyond some degree of randomness, as any other method. How is it robust against these fluctuations?

	Hereafter, we analyze the method performance in these real situations. Our strategy is simple: to test the method with clades for which one knows the entire past history since the first bifurcation. We construct these clades in a computer, following two ingredients \cite{Veit}. First, one starts from a single species. With a small fixed probability $b$, at each new time step the species can bifurcate. After that, each of the two emerging species evolves independent of the other. New branches may also bifurcate. One can book the exact times when each bifurcation occurs. An example of such a tree is shown in Figure 3.

	Second ingredient, each species' internal characteristics are represented by a sequence of $L$ bits 1 or 0. At each time step, this bitstring is mutated, i.e. an average number $m$ of its bits are randomly chosen and inverted from 0 to 1 or vice versa. When some species bifurcates, its current bitstring is copied to each new branch, both suffering independent random mutations thereafter. The scaling between $m$ and $L$ sets the maximum evolutionary time (time-back horizon) one can hope to reconstruct with the available accuracy. We use different seeds for random number generators governing the bifurcations ($R1$) or mutations ($R2$). Keeping the same $R1$ for different computer runs with different values of $m$, one can test the same tree topology under different mutation rates.

\begin{figure}[H]
 \begin{center}
 \input{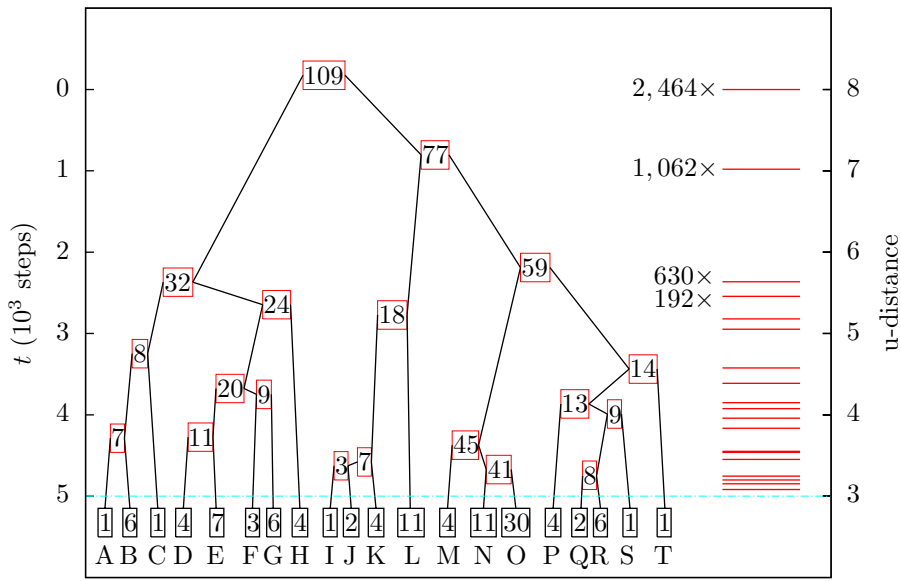}
 \end{center}
 \vskip-30pt
 \caption{Computer generated evolutionary tree. Starting from a single species, a first bifurcation occurs at $t=0$. A discrete clock $t=1,2,3\dots$ runs downwards. Each branch (species) can bifurcate with probability $b=0.0005$ at any time. When a bifurcation occurs, the number inside the corresponding square shows the quantity of alive species below it at $t=8,000$ (today, not shown) when one counts a total of $N=109$ alive species in this particular realization. Bifurcations occurred beyond $t=5,000$ are not shown for clarity.}

\label{simtree}
\end{figure}

	The distance between two bitstrings is the number of unmatched bits divided by $L/2$ (the random expected value) for normalization. Some similar normalization procedures are followed by linguists in order to cancel out phonetic accidental coincidences \cite{Wichmann}. The maximum  distance should be 1. However, due to fluctuations, old clades may present some distances slightly larger than 1, say a percentage $P\%$. They are of course statistically meaningless data. A nearly equal quantity below 1 is also supposed to be meaningless. Thus, $2P\%$ is a first, crude estimate for the dataset degree of distrust. Another approach would be the analysis of never-touched-bits, how this set still holding the original information shrinks as evolutionary time goes by \cite{Derrida}.

\begin{figure}[H]
 \vskip-70pt
 \begin{center}
 \input{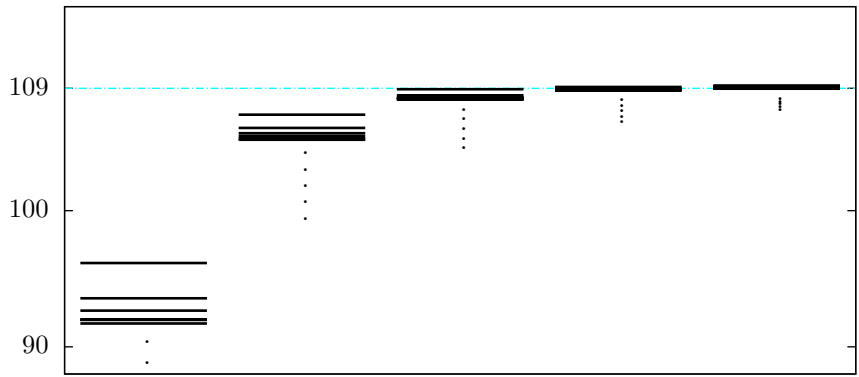}
 \end{center}
 \vskip-30pt
 \caption{Six topmost secular eigenvalues with increasing fluctuation ($m = 1.1$, $2.1$, $3.1$, $4.1$ and $5.1$ from left to right, $L=16,384$), for the same tree in Figure 3. Other 103 smaller eigenvalues are not shown (indicated by dots). For increasing fluctuation, the whole spectrum shrinks and saturates near the maximum conceivable value $N=109$ (except the always-present null Goldstone eigenvalue). Thus, in the limit of large fluctuations, the topmost eigenvalue becomes no longer isolated from the band below it, the genera partition may become wrong. In the current case this occurs at right, for $m=5.1$, when the $2,464$ distances corresponding to the highest level in Figure 3 form a wide band with only $16\%$ of the remainder $3,422$ data below it, as they all should be.}

\label{eigen}
\end{figure}

	Many clades like that exemplified in Figure 3 were tested, the result is indeed the expected one: perfect reconstruction up to a certain degree of randomness. The general behavior can be appreciated in Figure 4, showing (the top part of) the eigenvalue spectrum of matrix (5) --- not to be confounded with the spectrum of distances. At left, the reconstruction is perfect, while the highest eigenvalue is still separated from the band below it, thus corresponding to the correct partition eigenvector. Note also the saturation of the whole spectrum near the largest conceivable eigenvalue $N$, for increasing randomness. The correct partition is obtained up to $m=4.1$, where approximately $2P = 68\%$ of the whole dataset is statistically meaningless. For $m=5.1$ with $2P = 77\%$, the partition eigenvalue is surpassed by some other competitor, and the reconstruction fails. As a rule-of-thumb for real $D_{ij}$ datasets, they can be safely considered trustable if: I) The highest eigenvalue of matrix (5) is separated from the others; and II) At least the second largest eigenvalue is smaller than the upper limit $N$.

	In hard cases like the rightmost in Figure 4, the remote past of the first bifurcations is inaccessible with the accuracy at disposal. The measured dataset presents too much statistically meaningless entries (distances near $1$). They can be gradually expunged as follows. All $N(N-1)/2$ links between species are initially present, forming a completely connected network with $N$ vertices. One cuts the link corresponding to the largest distance $D_{ij}$, then the second largest, and so on. At some point along this sequence, the network becomes disconnected in two pieces. The very old history about this separation is inaccessible. But the recent history is not lost. The secular matrix method is then applied to each piece separately (expunged internal links re-included). If the above criteria I and II are not fulfilled for one piece, the cut-link procedure continues within it. Following this strategy for the tree in Figure 3 with $m=5.1$, the isolated species T is first disconnected from the other $108$. Then, also isolated species S disconnects, followed by A and by C. The set of $105$ remainder species still does not fulfill the correctness criteria I and II. Next, the block BH with $10$ species disconnects, and can be successfully reconstructed by the method. The remainder $95$ species do not. Continuing the cuttings, block IJK disconnects, then P, then FQR, then G and L, all successfully reconstructed, and so on.

	After each successful step where a correct bifurcation partition is found (under criteria I and II), the corresponding degeneracy is restored, changing the original dataset. Thus, one can re-start the whole process using the new dataset. Degeneracies are, thus, hierarchically and gradually restored. They can also be partially restored, by combining already defined blocks (groups among A, B, C $\dots$ R, S, T). After many runs, in case the complete reconstruction still fails, one can observe which blocks are most responsible for failures (normally isolated species or small sub-clades), and remove them from the dataset. Within the hard case $m=5.1$ in Figure 3, by removing isolated species A, C, S and T, and also blocks H, L and PQR, the remainder main tree with $N = 78$ species could be correctly reconstructed. Isolated, each removed block is also correctly reconstructed, but one cannot know where or when it should be branched from the main tree, because this occurred before the time-back horizon at disposal.

\section{Conclusion}

	There are a lot of alternative methods to reconstruct evolutionary trees (see, for instance, \cite{phylogeny,SaitouNei,Donetti,Capocci,Newman}). None of them considers the degeneracies appearing in the spectrum of real, ultrametric distances (past time since the last common ancestor), neither the breaking of these same degeneracies due to evolutionary drift, when measured through pairwise distances. This feature distinguishes the present method, the quoted degeneracies are gradually and hierarchically restored.

	The model considers {\it all} $N(N-1)/2$ distances at each step, mitigating the effects of the statistically meaningless part of the data. This feature is absent from traditional methods as the pioneering UPGMA, where the pair of species corresponding to the smallest distance is joined into a single species, whose distance to each remainder species is the average between both former distances. Instead of simply choosing the smallest distance, minimum-evolution approaches minimize quantities involving all distances, like our method, so improving the performance. Nevertheless, being yet neighbor-joining recipes, the number of distances is always reduced by $N-1$ at each step. Reference \cite{SaitouNei} presents such a method and comparisons with others. The best performances are equivalent to ours. Indeed, in many tests, whenever the correct partition is obtained by the method in \cite{SaitouNei}, it is also obtained by the current one. The tree in Figure 3, for instance, is correctly reconstructed up to $m = 4.1$ by both methods, but both fail under a little bit larger degree of randomness, $m = 4.2$. This coincidence indicates that reconstruction correctness is limited only by fluctuations in the measured dataset, not by drawbacks of the methods themselves. Moreover, when both fail, two different (wrong) partitions were observed. A posteriori, their comparison serves as a further criterion for reconstruction correctness.

	The current method belongs to the general class of spectral clustering for networks, see \cite{Donetti,Capocci,Newman}, based on the adjacency matrix $A_{ij} = 1$ or $0$ according to the edge between nodes $i$ and $j$ being present or absent. In our case all edges are present, but instead of $1$ or $0$ the corresponding matrix is constructed with the measured distances $D_{ij}$, real numbers. The same matrix (5) was heuristically adopted in \cite{Zhang}, without resorting to the current mechanical model. Once the eigenvector $\vec{x}$ corresponding to the highest eigenvalue $\lambda > 0$ is obtained, these authors adopted the following criterion: the $N$ elements of this vector are displayed in decreasing order, and the partition is defined where the largest gap between adjacent elements is found. Our not heuristic criterion, instead, is to take the partition according to the signs of the elements. The argument in favor of this criterion is straightforward: by solving Newton's law in Equation (3), $\ddot{\vec{x}} = \lambda \vec{x}$, one obtains $\vec{x}(t+\Delta t) \sim \exp{(\sqrt{\lambda} \Delta t)}\, \vec{x}(t)$. Thus, positive or negative elements of this vector exponentially grow in modulus as time goes by, and consequently the corresponding particles run away towards opposite senses along the $X$ axis. (That is why only the highest eigenvalue eigenvector, corresponding to the dominant value of $\lambda$, is responsible for the partition.) For small enough fluctuations in the measured distances, both criteria give the same result. Indeed, without fluctuations, the partition eigenvector is a completely flat step function, as in Figure 2, with a big gap separating positive from negative elements. By ``turning on'' the fluctuations, degeneracies and ultrametricity are broken, the step function bends towards a strictly decreasing monotonic behavior. The big gap remains but becomes smaller and smaller, up to the point where it no longer divides positive from negative elements. Moreover, this behavior serves as another, further criterion for reconstruction correctness: when the big gap separates elements of the same sign, the corresponding partition is not trustable. The numerical performance of the current method is the same as in \cite{Zhang}, the computer time required to find the eigenvector with the largest eigenvalue proportional to $N^2$.

\section{acknowledgments}

	The author is indebted to S{\o}ren Wichmann and Jos\'e Soares de Andrade Jr. for critical readings of the manuscript and helpful suggestions.

\end{document}